\documentstyle[12pt,epsfig]{article}
\newcommand{\ba}{\begin{eqnarray}}
\newcommand{\ea}{\end{eqnarray}}
\newcommand{\be}{\begin{equation}}
\newcommand{\ee}{\end{equation}}

\begin{document}
\begin{titlepage}
\begin{flushright}
CERN-TH/99-260\\
NTUA-74/99\\
hep-ph/9908462
 {\hskip.5cm}\\
\end{flushright}
\begin{centering}
\vspace{.3in}
{\bf  Unified Models at Intermediate Energy Scales}\\
{\bf and Kaluza--Klein Excitations }\\
\vspace{2 cm}
{G.K. Leontaris$^1$ and N.D. Tracas$^2$} \\ \vskip 1cm
{\it $^1$CERN  Theory Division, 1211 Geneva 23, Switzerland}\\
{ and}\\
{\it {Physics Department, University of Ioannina\\ Ioannina,
GR--45110, Greece}}\\
{\it {$^2$Physics Department, National Technical
University \\ 157 73 Zografou, Athens, Greece}}\\

\vspace{1.5cm} {\bf Abstract}\\
\end{centering}
\vspace{.1in}

We discuss the possibility of intermediate gauge coupling unification 
in unified models of string origin. Useful relations of the $\beta$-function
coefficients are derived, which ensure unification of couplings when 
Kaluza--Klein excitations are included above the compactification scale.
We apply  this procedure to two models with  $SU(3)\times SU(3)_L\times
SU(3)_R$ and $SU(4)\times O(4)$  gauge symmetries.

\vspace{2cm}
\begin{flushleft}
August 1999
\end{flushleft}
\hrule width 6.7cm \vskip.1mm{\small \small}
 \end{titlepage}

Recently, the possibility that the string and the compactification 
scale are around the energy determined by the geometric mean of the Planck mass
and the electroweak scale,  has appeared as a viable possibility in Type II 
string theories~\cite{IP}  with large extra dimensions~\cite{I}. 
On the other hand, as is well known, the minimal supersymmetric standard 
model (MSSM) spectrum leads to gauge coupling unification at a scale of 
$M_U\sim 10^{16}$GeV. To lower down this scale, usually 
power-law running of the gauge couplings is assumed, due to the
appearance of the Kaluza--Klein (KK) tower of states above the compactification
scale~\cite{D,R,B,LI,MQ}.

In a previous paper~\cite{lt}, we studied the possibility of intermediate
energy unification of the gauge couplings due solely to the presence of extra
matter and Higgs fields  under the standard  model (SM) group. We have found 
that unification may happen at the range $\sim 10^{11}$ GeV without the use of
power--law running from KK--excitations.  In this note we extend our analysis on
this issue by considering unified models of string origin which break down to the
SM group at some intermediate energy. We further assume the existence
of a compactification scale $M_C$ (smaller than the would be unification scale if
$M_C$ had not existed) above which KK--excitations are considered. In this context,
we find that unification can always be ensured whenever certain conditions of the
$\beta$-function differences are met.

We apply our results to models with intermediate gauge symmetries which 
involve no coloured gauge fields and can in principle be safe from 
proton decay operators. In particular, we study models based on the
$SU(3)^3$ and  $SU(4)\times O(4)$ gauge symmetries. Such models can
be derived from strings and possess various novel properties.
Among them, they possess particles with fractional charges
while they use small Higgs representations to break the gauge symmetry.
The superpotential possesses  various discrete and other
symmetries that may prevent undesired Yukawa couplings, while many
unwanted particles are projected out.  The original large gauge symmetry
breaks down to the intermediate gauge group of the type discussed above
owing to the existence of stringy type mechanisms. In the present
analysis we assume the existence of the  representations
that may be obtained in these models, and  the corresponding
KK--excitations. In our applications, below the intermediate breaking scale,
we assume the MSSM particle content, although our analysis can apply to 
any content respecting the general properties that we will derive in
what follows.

We start with the hierarchy of scales as they appear 
in our present work and which  are the following:
at the electroweak scale $M_W$,  we use  the  initial values for the gauge
couplings, as they are measured by the experiment. Next we consider
$M_S=1$ TeV, above which the MSSM $\beta$ functions are operative;
$M_G$, is the scale above which new physics appears and the 
$\beta$ functions of the specific grand unified model (GUT) are effective;
$M_C$, is the scale where compactification appears and the KK--states
start contributing to the $\beta$ functions, and 
$M_U$ denotes the scale where the gauge couplings would unify if there were no
compactification scale; $M_C$ is taken to be smaller than $M_U$.
 Finally, $M_{CU}$ is the scale where the gauge couplings unify when
we include the KK--excitations. We present them in  Fig.~1.

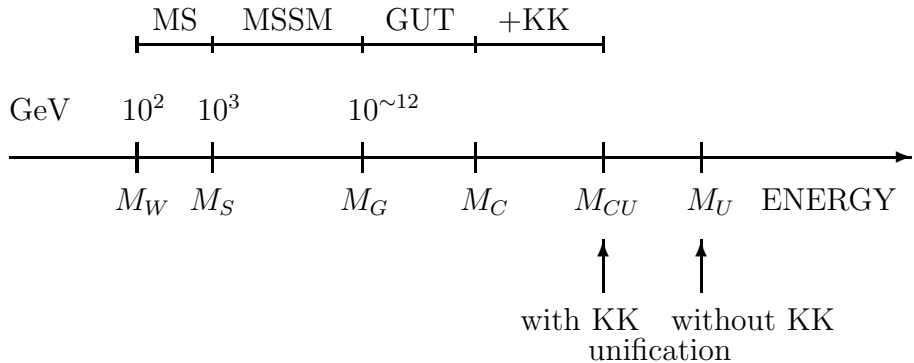
\begin{figure}
\setlength{\unitlength}{1mm}
\thicklines
\begin{picture}(130,70)

\put(5,40){\vector(2,0){120}}
\put(5,45){GeV}
\put(20,45){$10^2$}
\put(19,33){$M_W$}
\put(30,45){$10^3$}
\put(29,33){$M_S$}
\put(50,45){$10^{\sim 12}$}
\put(49,33){$M_G$}
\put(65,33){$M_C$}
\put(80,33){$M_{CU}$}
\put(95,33){$M_U$}
\put(105,33){ENERGY}

\put(22,55){\line(2,0){10}}
\put(22,38){\line(0,1){4}}
\put(22,54){\line(0,1){2}}
\put(24,57){MS}

\put(32,55){\line(2,0){20}}
\put(32,38){\line(0,1){4}}
\put(32,54){\line(0,1){2}}
\put(36,57){MSSM}

\put(52,55){\line(2,0){15}}
\put(52,38){\line(0,1){4}}
\put(52,54){\line(0,1){2}}
\put(55,57){GUT}

\put(67,55){\line(2,0){17}}
\put(67,38){\line(0,1){4}}
\put(67,54){\line(0,1){2}}
\put(70,57){+KK}

\put(84,38){\line(0,1){4}}
\put(84,54){\line(0,1){2}}
\put(97,38){\line(0,1){4}}

\put(84,22){\vector(0,2){7}}
\put(97,22){\vector(0,2){7}}

\put(73,17){with KK}
\put(93,17){without KK}
\put(82,13){unification}

\end{picture}
\begin{center}
\caption{The energy scales appearing in the paper}
\end{center}
\end{figure}

We begin our investigation along the lines discussed above, with the
presentation of a general property of the $\beta$--function coefficients.
Let $\beta_{ij}=\beta_i-\beta_j$ denote the $\beta$--function differences. 
We make the following two assumptions:
\noindent

\begin{itemize}
\item
There exists an energy scale $M_U$ where the coupling constants
$\alpha_i$'s unify, i.e.  $\alpha_i(M_U)=\alpha_U$ for all $i$,
assuming conventional logarithmic running (no--compactification scenario).
Quantitatively, this is expressed as
\begin{equation}
\frac{\alpha^{-1}_{ij}(M)}{\beta_{ij}}\equiv
\frac{\alpha^{-1}_j(M)-\alpha^{-1}_j(M)}{\beta_i-\beta_j}=
\frac{\alpha^{-1}_{ik}(M)}{\beta_{ik}}=
          \frac{1}{2\pi}\log\frac{M_U}{M}>0,
\label{cond1}
\end{equation}
where $M$ is some initial scale. The positiveness of the ratio ensures the
``convergence" (and not ``divergence") of the couplings above $M$. 
This point becomes essential when we discuss the cases of GUTs.
\item
The ratios of the differences of the $\beta$--functions $\beta^{KK}_{ij}$
(above the compactification scale $M_C$) to the corresponding
difference  $\beta^{ij}$ (below the compactification scale $M_C$) have the 
property:
\begin{equation}
\frac{\beta^{KK}_{ij}}{\beta_{ij}}=
\frac{\beta^{KK}_{ik}}{\beta_{ik}}>0.
\label{cond2}
\end{equation}
Again  positiveness ensures ``convergence" of the couplings above
$M_C$.
\end{itemize}
Then, it can be shown that the gauge couplings do unify, whatever energy
scale we choose as a compactification scale $M_C$, above which the
massive KK--states contribute to the running.

\noindent
Let us sketch the proof of the above statements~\cite{blois}.
Since all couplings unify at $M_U$ we have
\begin{equation}
\alpha^{-1}_U=\alpha^{-1}_i(M)-
              \frac{\beta_i}{2\pi}\log\frac{M_U}{M}.
\label{normalrunning}
\end{equation}
Assuming now that there exists a compactification scale $M_C<M_U$,
the running of the couplings, for $M'>M_C$, is given by
\footnote{ We ignore the contribution of the MSSM massless states above $M_C$
since it is negligible compared to that of the KK--excitations.
We use the successful approximation of incorporating the massive KK--states
with masses less than the running scale~\cite{D}.} 
\begin{equation}
\alpha^{-1}_i(M')=\alpha^{-1}_i(M_C)-
              \frac{\beta^{KK}_i}{2\pi}\left(
2N\log\frac{M'}{M_C}-2\log(N!)\right).
\label{basic}
\end{equation}
where $N$ is an integer such that $(N+1) M_C > M' > N M_C$, which counts
the massive KK--states that have masses below the running scale (we have
assumed only one extra dimension and in that case the multiplicity of the
states at  each mass level is 2). From the running below $M_C$, we can
express $\alpha^{-1}(M_C)$ in the form
\[
\alpha^{-1}_i(M_C)= \alpha^{-1}_i(M)-
              \frac{\beta_i}{2\pi}\log\frac{M_C}{M}=
              \alpha^{-1}_U-\frac{\beta_i}{2\pi}\log\frac{M_C}{M_U}
\]
and (\ref{basic}) is written as
\begin{equation}
\alpha^{-1}_i(M')=\alpha^{-1}_U-\frac{\beta_i}{2\pi}\log\frac{M_C}{M_U}-
              \frac{\beta^{KK}_i}{2\pi}\left(
2N\log\frac{M'}{M_C}-2\log(N!)\right).
\label{basicp}
\end{equation}
Suppose now that the two couplings $\alpha_i$ and $\alpha_j$ meet at the
energy scale $M_{CU}$. It is easy to check that the following
relations hold:
\begin{eqnarray}
2N\log\frac{M_{CU}}{M_C}-2\log(N!)&=&-\frac{\beta_{ij}}{\beta^{KK}_{ij}}
\log\frac{M_C}{M_U}\nonumber\\
\alpha^{-1}_i(M_{CU})=\alpha^{-1}_j(M_{CU})&=&
     \alpha^{-1}_U-\frac{\beta_i}{2\pi}\log\frac{M_C}{M_U}+
\frac{\beta^{KK}_i}{2\pi}
\frac{\beta_{ij}}{\beta^{KK}_{ij}}\log\frac{M_C}{M_U}\nonumber\\
&=&\alpha^{-1}_U-\frac{\beta_j}{2\pi}\log\frac{M_C}{M_U}+
\frac{\beta^{KK}_j}{2\pi}\frac{\beta_{ij}}{\beta^{KK}_{ij}}\log\frac{M_C}{M_U}.
\label{basicpp}
\end{eqnarray}
The value of the third coupling $\alpha^{-1}_k(M_{CU})$ at the scale
$M_{CU}$ is given by
\begin{eqnarray}
\alpha^{-1}_k(M_{CU})&=&
     \alpha^{-1}_U-\frac{\beta_k}{2\pi}\log\frac{M_C}{M_U}-
              \frac{\beta^{KK}_k}{2\pi}\left(
2N\log\frac{M_C}{M'}-2\log(N!)\right)
              \nonumber\\
&=& \alpha^{-1}_U-\frac{\beta_k}{2\pi}\log\frac{M_C}{M_U}+
              \frac{\beta^{KK}_k}{2\pi}
               \frac{\beta_{ij}}{\beta^{KK}_{ij}}\log\frac{M_C}{M_U}.
\end{eqnarray}              
It is now straightforward to check, using the second condition
(\ref{cond2}), that $\alpha^{-1}_k(M_{CU})$ equals the values of
$\alpha^{-1}_i$ and $\alpha^{-1}_j$ at the same scale. Therefore, the
couplings unify, no matter what compactification scale $M_C$ we choose. 
The positiveness condition of (\ref{cond2}) comes from the ``convergence''
requirements of the couplings above $M_C$. From (\ref{basic}) we get 
\[
\frac{\alpha^{-1}_{ij}(M_C)}{\beta^{KK}_{ij}}=\frac{1}{2\pi}
                          \left( 2N\log\frac{M_{CU}}{M_C}-2\log(N!)\right),
\]
which should be positive, since the unification scale $M_{CU}>NM_C$. 
But from the running below $M_C$ we get
\[
\alpha^{-1}_{ij}(M_C)=\frac{\beta_{ij}}{2\pi}\log\frac{M_C}{M}
\]                          
and the positivity condition can be put in the form
\[
\frac{\beta_{ij}}{\beta^{KK}_{ij}}>0.
\] 

Let us note also that the initial scale $M$ in (\ref{cond1})
could be either an intermediate one where a group larger than the SM one 
appears, or could be just $M_W$ if no GUT is assumed.

We now come to the $\beta$--function, both below and
above $M_C$. Below the compactification scale, the (one--loop)
$\beta$--function is given by
\begin{equation}
\frac{1}{16\pi^2}\left( -3C_2(G)+\sum_i T(R_i)\right),
\label{bfunction}
\end{equation}
where the first term corresponds to the vector supermultiplet 
(gauge bosons and gauginos) contribution while,
 the second corresponds to the chiral
(quarks, leptons, higgs and superpartners) supermultiplets. 
$C_2(G)$ is the quadratic Casimir operator for the adjoint representation,
$R_i$ is the representations of the matter multiplets and $T(R)$ is
defined by the relation $Tr[R^aR^b]=T(R)\delta^{ab}$.
Above $M_C$, the massive KK--states
give the following $\beta$--function
\begin{equation}
\frac{1}{16\pi^2}\left( -2C_2(G)+\sum_i T(R_i)\right).
\label{KKbfunction}
\end{equation}
The difference from (\ref{bfunction}) comes from the fact that the
massive vector supermultiplet is actually a $N=2$ hypermultiplet with 
a vector plus a chiral supermultiplet. 

As a first example we discuss the MSSM where we know that the three
couplings $\alpha_1$, $\alpha_2$ and $\alpha_3$ unify at the scale
$\sim 10^{16}$GeV. Now assuming that only the gauge bosons and the higgs
acquire KK--states (the matter fields are placed on the fixed points
of the heterotic string and therefore no KK--states appear for them),
the above formulae give
\begin{equation}
\begin{array}{lll}
16\pi^2 \beta_{31}=-9.6, &
16\pi^2 \beta_{32}=-4,   &
16\pi^2 \beta_{21}=-6.4,\\
16\pi^2 \beta^{KK}_{31}=-6.6, &
16\pi^2 \beta^{KK}_{32}=-3,   &
16\pi^2 \beta^{KK}_{21}=-4.4
\end{array}
\end{equation} 
Therefore, with an error of less than 10\%, the
ratio $\beta_{ij}/\beta_{ik}$ is the same below and above $M_C$. Note here
that, since the matter multiplets are complete $SU(5)$ ones (the equal
contribution of matter in the three $\beta$-functions is due to that),
even in the case where they had  KK--excitations, the relations between the
$\beta$-function ratio would still hold. Therefore, whatever energy scale
we choose as our compactification scale, the three couplings will unify.
We now  apply this idea to the two models mentioned above. Some 
details on the $\beta$-functions and the string spectra of the
models may be found in~\cite{alt}.

{\it{\underline{The $SU(4)\times O(4)$ case}}}

 We first 
take as an example the $SU(4)\times SU(2)_L\times SU(2)_R$ model, which
is assumed to break to the SM--symmetry at some scale $M_{G}$. Above
$M_{G}$, apart from the MSSM matter content, we have the following
extra states
\begin{eqnarray}
n_6=(6,1,1),\quad n_4=(4,1,1),&\quad& n_L=(1,2,1),\quad n_R=(1,1,2)
\nonumber\\
n_{22}=(1,2,2),&\quad& n_H=(4,1,2)/({\bar 4},1,2).\nonumber
\end{eqnarray}    
where we show the quantum numbers under the GUT group. The subscript $H$ 
refers to the Higgs fields that break the $SU(4)$ and the $SU(2)_R$ groups, 
while the 22 gives the Standard Model Higgs. The one loop $\beta$--functions
are
\begin{eqnarray}
\beta_R&=&-6+2n_G+2n_H+2n_{22}+n_R/2\label{bR}\nonumber\\
\beta_L&=&-6+2n_G+2n_{22}+n_L/2\label{bL}\label{b4}\\
\beta_4&=&-12+2n_G+n_H+n_6+n_4/2.\nonumber
\end{eqnarray}
where $n_G$ is the number of generations. The relations between
the MSSM and the GUT model couplings, at $M_G$, are
\[
\alpha_4=\alpha_3, \quad \alpha_L=\alpha_2, \quad
\alpha^{-1}_R=(5/3)\alpha^{-1}_1-(2/3)\alpha^{-1}_4.
\]
Assuming now that the 
``turning'' point from MSSM to the GUT content is $10^{11-14}$ GeV,
the ratios of the coupling constant differences are in the ranges
\[
\frac{\alpha^{-1}_{4R}}{\alpha^{-1}_{4L}}=3.54-3.79,\quad
 \frac{\alpha^{-1}_{L4}}{\alpha^{-1}_{LR}}=(-0.39)- (-0.36).
\]
Above the compactification scale we assume that all extra (beyond
that of the MSSM) matter could have KK--states.
Allowing  a difference at most 3\% between the ratio of the
coupling constants and the ratio of the $\beta$--functions, and
for $M_G=10^{12}$ GeV and $M_G=10^{13}$ GeV, the only 
values that the $\beta$-function can give (all $n's$ take even integer
values) are
\[
\frac{\beta_{4R}}{\beta_{4L}}=\frac{\beta^{KK}_{4R}}{\beta^{KK}_{4L}}
=\frac{33}{9},\quad
 \frac{\beta_{L4}}{\beta_{LR}}=\frac{\beta^{KK}_{L4}}{\beta^{KK}_{LR}}=
-\frac{3}{8}.
\]
If we require $M_G$ to be either $10^{11}$ GeV or $10^{14}$ GeV,
then we should raise the acceptable error between the ratios to 5\%
and the only values that the ratios, below $M_C$, can have are
\[
\frac{\beta_{4R}}{\beta_{4L}}=\frac{33}{9}{\mbox{ or }}\frac{7}{2},\quad
 \frac{\beta_{L4}}{\beta_{LR}}=-\frac{3}{8}{\mbox{ or }}-\frac{2}{5},
\]
while the ratios  above $M_C$ remain the same. 

Of course, several particle contents below and above the compactification
scale,
render the above values for the ratios. In the following table
we give one example, where the content below $M_C$ can, in principle, be
reproduced by the string $SU(4)\times SU(2)_L\times SU(2)_R$ model,
while we have chosen $M_G=10^{12}$GeV
\begin{equation}
\begin{array}{lcccccc}
                   & n_6   & n_4  & n_L & n_R &  n_H  &  n_{22}\\
{\mbox{ below }M_C}& 4     &   8  & 10  & 10  &  4    &    4\\
{\mbox{ above }M_C}& 0     &   2  &  0  &  0  &  4    &    4\,.
\end{array}
\label{422content}
\end{equation}
In Fig.~2 we show the running of the coupling constants for the above
content and for several values of $M_C$. In Fig.~3  a scatter plot is
presented 
showing the (inverse) of the unified coupling for several contents of
the model.
\begin{figure}
\begin{center}
\hspace{-1cm}
\epsfig{figure=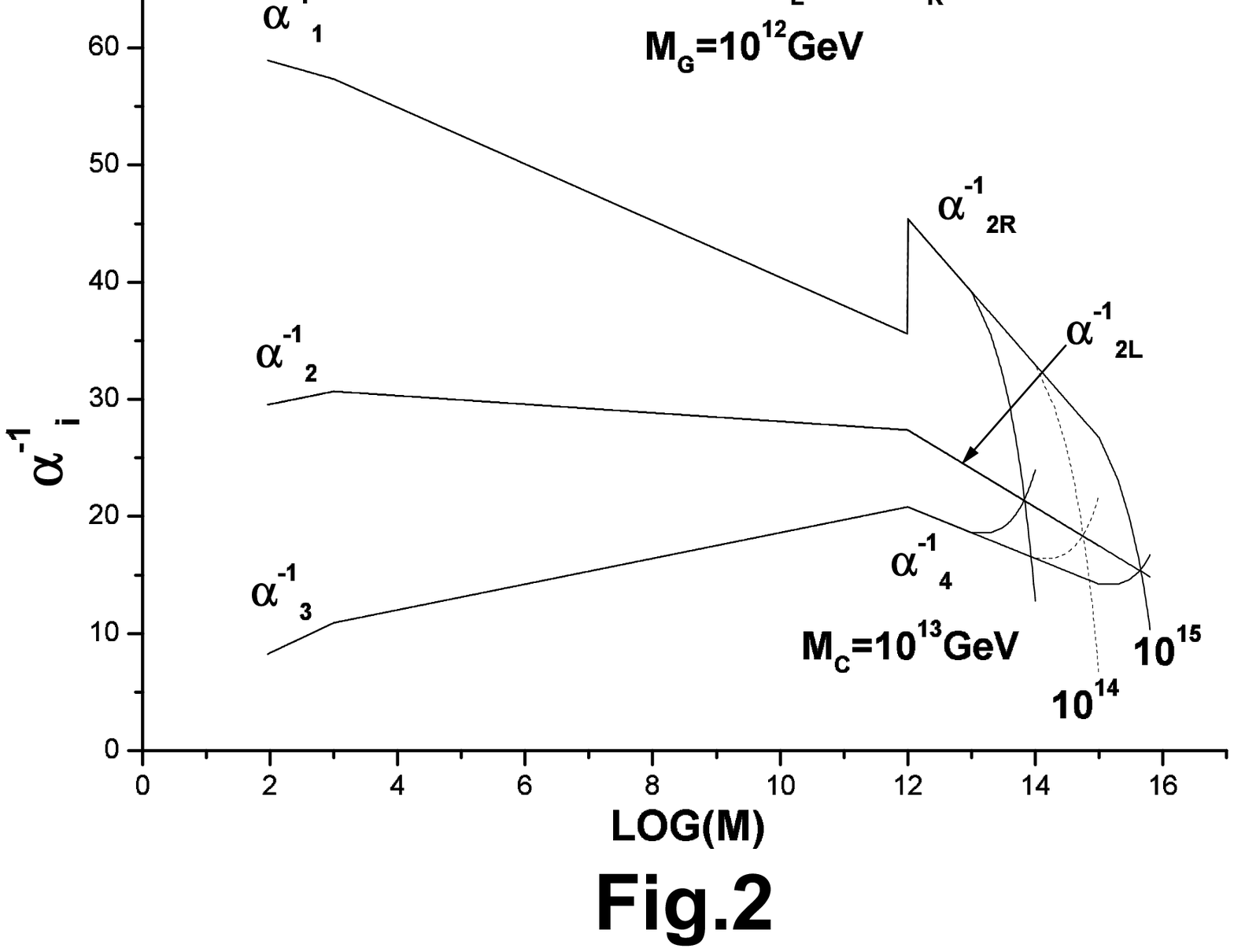,width=12cm}
\vspace{.5cm}
\caption{The inverse of the three gauge couplings as a function
of energy, for the $SU(4)\times SU(2)_L\times SU(2)_R$
GUT with the specific content appearing in
(\ref{422content}). 
We have chosen $M_G=10^{12}$GeV and three values of the
compactification scale $M_C=10^{13},10^{14},10^{15}$GeV.
}
\end{center}
\label{fff2}
\end{figure}
\begin{figure}
\begin{center}
\hspace{-1cm}
\epsfig{figure=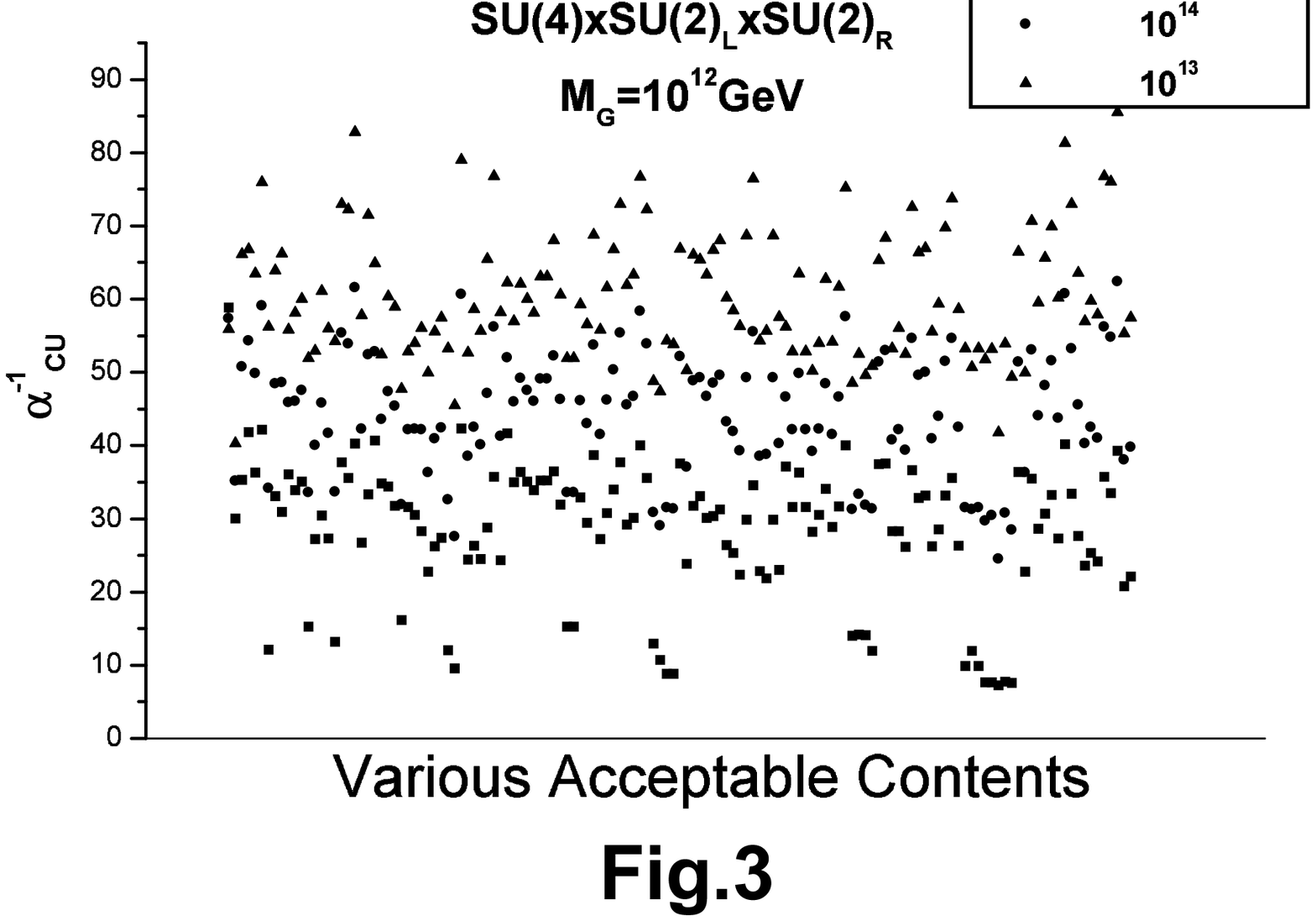,width=12cm}
\vspace{.5cm}
\caption{Scatter plot of the inverse of the unified gauge coupling,
for the $SU(4)\times SU(2)_L\times SU(2)_R$
GUT, choosing $M_G=10^{12}$GeV  for three values of the
compactification scale $M_C=10^{13},10^{14},10^{15}$GeV.
The horizontal axes  enumerates the various acceptable
contents of the model (the order of appearance along the $x$--axis is irrelevant). 
The highest the compactification scale
the greater the value of the $\alpha_{CU}$.}\end{center}
\label{fff3}
\end{figure}

{\it{\underline{The $SU(3)_C\times U(3)_L\times U(3)_R$ model}}}

Another interesting string derived model, which  admits a low (intermediate)
unification scale (no dangerous dimension--six operators), is based on the
$SU(3)\times SU(3)_L\times SU(3)_R$ symmetry. The MSSM content
is found in the $27$ representation of the $E_6$ group
\begin{equation}
27\rightarrow (3,{\bar 3},1)+({\bar 3},1,3)+(1,3,{\bar3}).
\label{normal}
\end{equation}
where
\begin{equation}
(3,{\bar 3},1)=\left(
\begin{array}{c}u\\d\\D\end{array}\right),
({\bar 3},1,3)=\left(
\begin{array}{c}u^c\\d^c\\D^c\end{array}\right),
(1,3,{\bar3})=\left(
\begin{array}{ccc}
h^0&h^+&e^c\\
h^-&{\bar h}^0&\nu^c\\
e&\nu&N
\end{array}\right).
\end{equation}

The breaking chain  we adopt here is the following: the first
group is the colour $SU(3)$. The second breaks to $SU(2)_L\times U(1)_L$,
 while the third breaks to a $U(1)_R$. The SM $U(1)_Y$ emerges as a
linear combination of the two $U(1)_{L,R}$. The conventional hypercharge
$Y$ is related to the $X$ and $Z$ charges of $U(1)_L$ and $U(1)_R$ 
correspondingly, by the relation
\[
Y=\frac{1}{\sqrt 5}X+\frac{2}{\sqrt5}Z,
\]
while the corresponding relations of the couplings at the breaking scale
is
\[
\alpha_L=\alpha_2,\quad
\alpha^{-1}_R=(5/4)\alpha^{-1}_{Y}-(1/4)\alpha^{-1}_{L}.
\]

Apart from the above states, in the string model, fractionally charged
 and other exotic states usually appear,
belonging to the representations
\begin{equation}
\begin{array}{|c|c|c|}
\hline
(\stackrel{(-)}{3},1,1) & (1,\stackrel{(-)}{3},1) &
(1,1,\stackrel{(-)}{3})\\
\hline
0  &   \pm 1/3 {\mbox{ and}}\pm 2/3 & \pm 1/3 {\mbox{ and}}\pm 2/3\\
\hline
\end{array}
\label{exotic}
\end{equation}
where the second line shows the corresponding (electric) charges.
One should not be misled by the values of these charges: the neutral
states are coloured, while the others are singlet under the colour group.
Therefore, after the symmetry breaking, these states will result in exotic 
lepton doublets and singlets carrying charges like those of the down and up
quarks. Note that such states are not common in GUTs, however, they are
generic in string models.

The one--oop $\beta$--functions are given by
\begin{eqnarray}
\beta_3&=&-9+\frac{1}{2}\left(3n_Q+3n_{Q^c}+n_C\right)\\
\beta_L&=&-9+\frac{1}{2}\left(3n_Q+3n_L+n_{L'}\right)\\
\beta_R&=&-9+\frac{1}{2}\left(3n_{Q^c}+3n_L+n_{L''}\right),
\end{eqnarray}
where $n_Q$, $n_{Q^c}$ and $n_L$ are the number of the representations
appearing in the complete 27, Eq.~(\ref{normal}), while $n_C$, $n_{L'}$
and $n_{L''}$ are the number of the exotic representations of
(\ref{exotic}).

As in the case of the previous  model, several massless spectra  pass
the two conditions and provide unification of the three couplings. Although it
seems that the $SU(3)^3$ is probably less constrained (giving a lot of possible
contents, presumably because of the symmetric form of the $\beta$-functions),
one should be careful, since the unification coupling could be high enough
in  some cases and get out of the perturbative region. This of course happens
for high matter content, when the $\beta$--functions become large and positive.
We should note at this point (and it is a general remark not applicable only to
the specific GUT) that the value of $M_C$ starts playing a significant
role in the case where the unification coupling constant is getting large: if the
$\beta$-functions between $M_G$ and $M_C$ are already large, $M_C$ cannot be much
larger than $M_G$ if we want to avoid a non-perturbative value of the unification
coupling.

In the following table, we give, as an example, the content below and
above $M_C$, for the $SU(3)^3$ model, where we have chosen $M_G=10^{12}$ GeV  
and  a 3\% error in the equality of the ratios
\begin{equation}
\begin{array}{lcccccc}
                   & n_Q   & n_L  & n_{Q^c} & n_C &  n_{L'}  &  n_{L''}\\
{\mbox{ below }M_C}& 10    &   4  &  6      & 10  &  0       &    2\\
{\mbox{ above }M_C}& 6     &   0  &  2      & 10  &  0       &    2\,.
\end{array}
\label{333content}
\end{equation}
In Fig.~4 we show the running of the couplings for the above content and
for  several values of $M_C$ while Fig.~5 is a scatter plot of the (inverse of the)
inified coupling foe several contents of the model.

\begin{figure}
\begin{center}
\hspace{-1cm}
\epsfig{figure=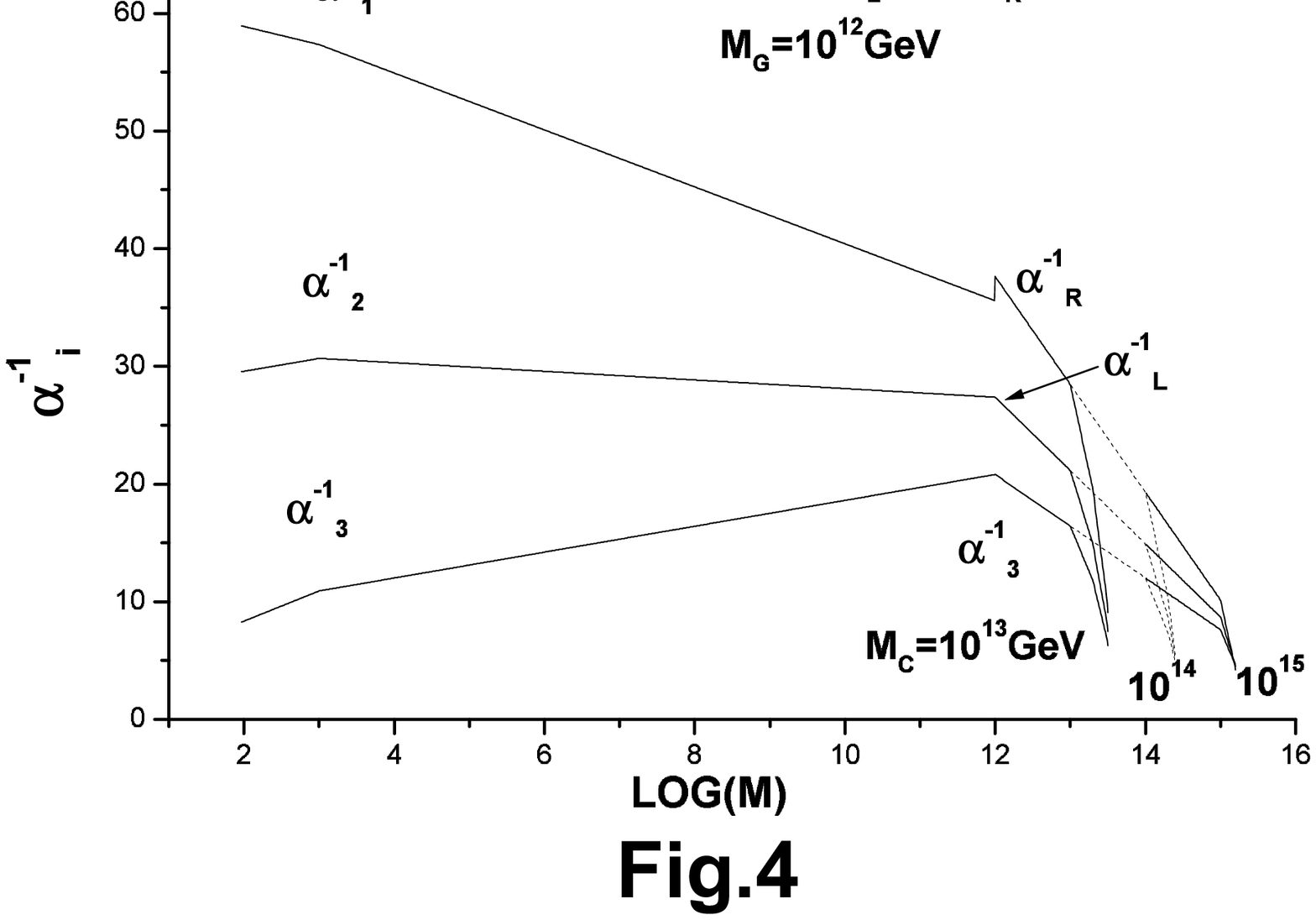,width=12cm}
\caption{
Same as in Fig.2 for the $SU(3)\times SU(3)_L\times SU(3)_R$  model and the specific 
content of (\ref{333content}).
}
\end{center}
\end{figure}

\begin{figure}
\begin{center}
\hspace{-1cm}
\epsfig{
figure=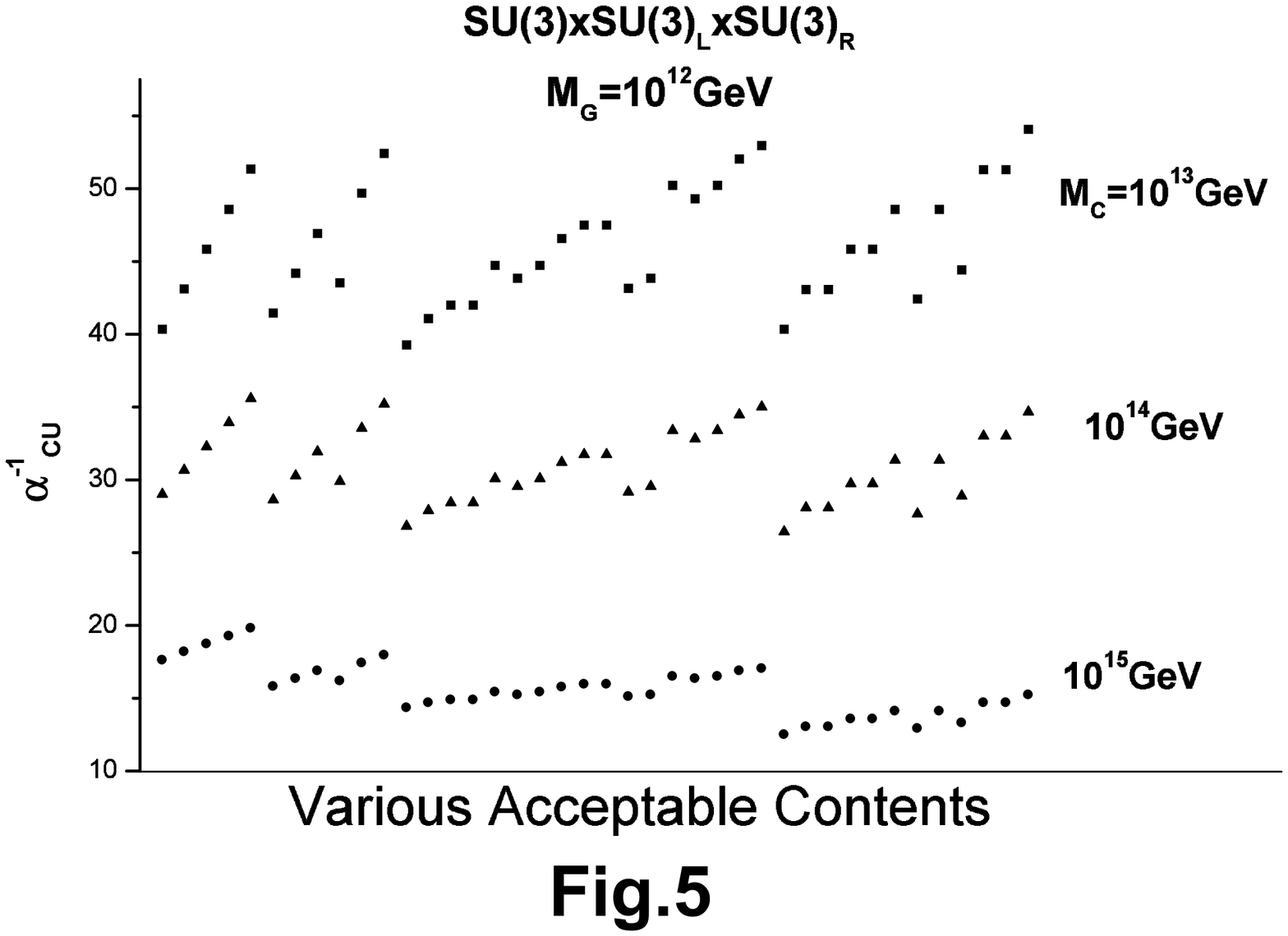,width=12cm}
\caption{
Same as in Fig.3 for the $SU(3)\times SU(3)_L\times SU(3)_R$ model.}
\end{center}
\end{figure}

We conclude with a few remarks: the possibility of lowering 
the unification scale is a fascinating one, both from the theoretical
and from the experimental point of view. Experimentally, it would 
be exciting to have a low enough unification scale for 
the possibility of testing its implications in the near--future machines.
Theoretically, it would give a solution to the desert-puzzle invoked in
previous Planck--mass unification scenarios. However, when lowering the unification
scale in most of the GUTs, one faces the notorious problem  of proton
decay.  A possible solution, which combines the idea of a relatively low
unification and a reasonable solution to the proton decay problem, is
the one presented in this note. We have considered GUTs that do not
lead to proton decay via dimension-six operators and 
implemented the idea that the unification occurs at an intermediate
scale so that, for appropriate  Yukawa couplings, other dangerous operators  
may be sufficiently suppressed. We have shown that there exist numerous
cases of massless spectra (which can  be derived from the superstring),
implying naturally intermediate scale unification.

\end{document}